\documentclass[aps,prl,twocolumn,showpacs]{revtex4}
\usepackage{graphicx}
\begin{document}
\title{Finite hadronization time and unitarity in quark recombination model}
\author{C.B. Yang}
\affiliation{Institute of Particle Physics, Central China Normal University,
Wuhan 430079, P.R. China}
\date{\today}
\begin{abstract}
The effect of finite hadronization time is considered in the recombination model,
and it is shown that the hadron multiplicity turns out to be proportional to
the initial quark number and unitarity is conserved in the model. The baryon to
meson ratio increases rapidly with the initial quark density due to competition
among different channels.

\end{abstract}

\maketitle

In last few years, the quark recombination (or coalescence) model \cite{reco,mv,anom,fb}
has aroused intense interest in the communities of both theorists and
experimentalists in the field of ultra-relativistic heavy ion collision physics,
because of its ability to explain novel phenomena, such as unexpected high
proton over pion ratio at $p_T$ about 3 GeV$/c$ \cite{popi} and the constituent
quark number scaling of the elliptic flow \cite{v2}, that are observed in RHIC experiments.
Although different physics considerations are input in the investigation of
final state particle spectra in different implementations of the quark recombination
model by different groups, there are a lot of common points \cite{fri}.
All implementation of the model, except the latest attempt \cite{rec3}, considered
only quark and antiquark degrees of freedom and gluons are assumed having been
converted into quarks and antiquarks before hadronization. Every group assumed
implicitly that hadronization occurs instantly. In heavy ion collision processes
there are plenty of soft thermalized partons and some hard partons which will evolve into
showers of semihard partons \cite{hy1}. Because of the absence of gluon degree of freedom
in the model, quarks (and antiquarks) at the moment just before hadronization
must have been dressed up as constituent quarks ready for recombining into final
state hadrons. It has been shown that the recombination of
thermal quarks dominates in Au+Au collisions at RHIC energies
for low $p_T$ pion production and the traditional fragmentation,
which can be interpreted as a recombination process from shower partons originated
from a hard parton \cite{hy1}, dominates at very high $p_T$, while the thermal-shower
recombination is important for moderately high $p_T$, $3{\rm GeV}/c<p_T<8{\rm GeV}/c$
in central Au+Au collisions \cite{hy2}. In this way, all hadron production can be considered
consistently in the framework of quark recombination model.
In every implementation of the quark recombination model, the transverse
momentum spectrum of mesons at mid-rapidity can be written, after some algebra, as
\begin{equation}
\frac{dN^M}{p_Tdp_T}=\int dp_1dp_2F(p_1,p_2)R^M(p_1,p_2,p_T)\ ,
\label{old}
\end{equation}
where a factor $\delta(p_T-p_1-p_2)$ is included in the recombination
function $R^M(p_1,p_2,p_T)$ to ensure momentum conservation and $F(p_1,p_2)$
is associated with the joint quark-antiquark momentum distribution.
When we are interested in the total
multiplicity of a kind of meson, we can consider the contribution from
thermal-thermal recombination only, because most produced hadrons are in the
low transverse momentum region where pure thermal recombination
dominates. In this range of transverse momentum, the joint distribution for
quark-antiquark pair can be written as $F(p_1,p_2)=V\rho^2f_1(p_1)f_2(p_2)$
with $V$ the spatial volume of the partonic system and $\rho$ the thermal
parton density just before hadronization when the (anti)quark transverse
momentum distributions $f_{1,2}$ are normalized to 1.
A simple conclusion from Eq. (\ref{old}) is that the total
yield of the meson is proportional to the square of quark density just before
hadronization. Similarly, one can conclude that the yield of a kind of baryon
should be proportional to the cubic of the quark density. On the one hand, if
one considers production of pentaquarks along the same line, one has to say that
its yield is proportional to $\rho^5$ and higher in heavy ion collisions than
in elementary ones. This conclusion seems to contradict to current experimental
observations. On the other hand, that the yield in a given coalescence channel scales
quadratically/cubically with the constituent number violates unitarity, as argued in
\cite{mv,mol,lm}, because the total number of hadrons should be linearly proportional
to that of constituent quarks just before hadronization. Apparently, the total
number of constituent quarks in final state hadrons should be equal to the
quark number just before hadronization. Considering above, one has to conclude
that something must be changed in current formulism of the recombination model.

An important observation in this paper is that the numbers of constituent
quarks are decreasing at finite rate during the hadronization process if
one assumes that the hadronization lasts a nonzero period of time. Then
one should study the effect
of finite hadronization time on the yields of hadrons in the recombination model.
For this purpose, one can interpret the recombination function $R^M$ as the average
production rate of meson from a given pair of quark and antiquark
and write the joint two-quark distribution as a function of time explicitly.
Then the right hand side of Eq. (\ref{old}) should be $dN^M/p_Tdp_T dt$. Here
time $t$ can be defined in the laboratory frame.
In this paper, we will show that the introduction
of the finite hadronization time in the recombination model can ensure unitarity.

Instead of considering the spectra of different hadrons in the framework of quark
recombination models, we only investigate the yields of mesons and baryons, as done
in earlier work of the coalescence model \cite{coals}. For simplicity, we assume
in this paper that quarks and antiquarks have the same density in the almost baryon
free central rapidity region and that the shapes of the all transverse momentum
distributions involved remain unchanged in the hadronization process. The strange contents
are not considered in this paper. If the phase transition from partons to hadrons
is of first-order, there is a boundary and hadrons are produced only on the surface
while the partons inside have a constant temperature and/or density. If, on the other
hand, the phase transition is of second-order, no boundary between parton state and hadron
state can be defined, and hadronization takes place in the whole system at the same time.
Most event generators for high energy collisions adopt the second scenario in the codes.
Such a scenario was also taken in \cite{biro99}. In ultra-relativistic heavy ion collisions
the produced partonic system expands. Then the phase transition may be neither first-order
nor second-order. If the expansion retains to the last stage of the evolution and the
phase transition is likely first-order, there are two competing trends on the change of the
volume. The expansion will make the partonic volume larger, and the hadronization tries
to shrink the system. Thus the hadronization dynamics should, generally speaking, be
connected with the hydrodynamical calculations. This cannot be included in this paper.
Now we simply consider two limiting cases for the evolution of the volume in hadronization
process.

The first case we consider is for fixed volume for partonic system. In other words,
the expansion of the system is assumed to be compensated by the shrinking from
hadronization. Because the hadronization time is very short, this case can be regarded
as the limit of extremely rapid expansion.

First of all, we assume that only one species of meson can be produced in the process.
Then one can integrate over the momentum of produced hadrons and get the production
rate of the meson as
\begin{equation}
{dN^M\over dt}=AV\rho^2(t)\ ,
\end{equation}
and the information of both the shape of quark distribution and the recombination
function is encoded in $A$. No reverse term for the process
from hadron to quarks is considered here,  because (i) most of the
produced hadrons will move quickly to the detectors, thus the hadron
density in the reaction zone is much smaller than that for quarks, and
(ii) at the energy scale corresponding to the phase transition (about 170 MeV),
it is very unlikely for hadrons to dissolve into quarks and antiquarks in
the interactions. As a
consequence, the interactions among the produced hadrons will be elastic and no
lose term appear in Eq. (2).
With the production of the meson, the constituent
quark number decreases accordingly, and we have
\begin{equation}
\frac{d(V\rho)}{dt}=-\frac{dN^M}{dt}\ .
\end{equation}
This equation comes from the model assumption that just before hadronization all partons
have been converted into constituent quarks ready for recombination into final
state hadrons and no more constituent quarks can be created. Finally, all those
$N_q+N_{\bar{q}}$ constituent quarks have to be inside some hadrons, because of
color confinement.
Thus as time goes to infinity, $\rho$ must go to zero from initial density $\rho_0=N_q/V$
just before hadronization. Equations corresponding to Eqs. (2) and (3)
can be found for any chemical reaction $A+B\to C$ which is obviously a combination
process similar to the process $q+\bar{q}\to \pi$ studied in the quark recombination
model. For chemical reactions the amount of final
product always depends linearly on the initial density. From above two equations,
one can easily deduce the yield as
\begin{equation}
N^M=V\rho_0=N_q\ .
\label{mul}
\end{equation}
Therefore, the total multiplicity of the meson turns out to be equal to
the initial number of constituent quark, and unitarity is maintained in
the recombination process. A careful reader may have noticed that the parameter $A$
introduced in Eq. (2) is not relevant to the total multiplicity of the meson, as shown
in Eq. (\ref{mul}). This is expected from quark number counting.

If only one species of hadrons consisting of $n$ constituent quarks is considered,
the total multiplicity $N^n$ can be shown, along the same line as for mesons, to be
\begin{equation}
N^n=\frac{2V\rho_0}{n}=\frac{2N_q}{n}\ .
\end{equation}
Again the total multiplicity is proportional to the initial number of constituent quark
and there is no violation of unitarity.

In real high energy collisions, many different species of particles are
produced. In the recombination model, the production rates for different hadrons depend on
quark distribution differently. For mesons the production rate depends on quark density
quadratically, while for baryons the dependence is of cubic. So
there exist competitions among different channels and the effect of competition depends on
the initial number of quarks. To get a parameter free
expression, we assume in this investigation that only non-strange mesons and
(anti)baryons are produced. Then we have
\begin{eqnarray}
&{dN^M\over dt} &=A_1V\rho^2(t)\ ,\nonumber\\
&\frac{dN^B}{dt} & = A_2V\rho^3(t)\ ,\label{basic}\\
&\frac{d(V\rho)}{dt} &=-\left(\frac{dN^M}{dt}+3\frac{dN^B}{dt}\right)\ ,\nonumber
\end{eqnarray}
with initial conditions $N^{M,B}=0$ at $t=0$, $\rho(0)=\rho_0$. There are exactly
the same number $N^B$ anti-baryons. The third expression in last equation ensures
the unitarity in hadronization process. Because unitarity condition was not
taken into account in former applications of the recombination model, the result
in this paper will not be reduced to the old ones even in the limit of zero
hadronization time. From the reaction-rate theory \cite{react}
$A_1$ and $A_2$ are proportional to the corresponding cross-sections
and the degeneracy factors. Above equations
can be solved numerically. First we can define dimensionless quantities $u=3A_2\rho_0/A_1,
r=\rho(t)/\rho_0, \tau=A_1\rho_0t$.
Here variable $u$ has encoded initial quark density and the competition between meson
and baryon channels. With these scaled variables above equations can be normalized to
\begin{eqnarray}
\label{nm}
  \frac{dN^M}{ud\tau} &=& \frac{VA_1}{3A_2}r^2(u,\tau) \ ,\\
  \label{nb}
  \frac{dN^B}{ud\tau} &=&\frac{VA_1}{9A_2}ur^3(u,\tau)\ ,
  \end{eqnarray}
where the scaled quark density $r(u,\tau)$ at scaled time $\tau$ is the
solution of the equation
  \begin{equation}
  \tau = u\ln r+\frac{1}{r}-1-u\ln\frac{1+ur}{1+u}\ .
  \label{evol}
\end{equation}

\begin{figure}[tbph]
\includegraphics[width=0.45\textwidth]{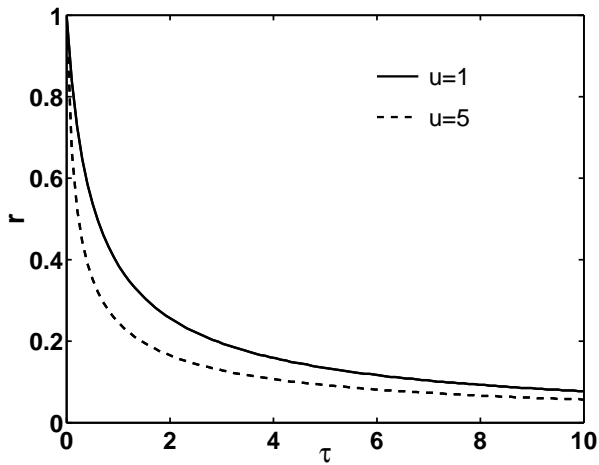}
\caption{Behavior of scaled density as a function of scaled
time for fixed $u=1$ and 5.}
\end{figure}

The relative density $r$ are shown in Fig. 1 as a function of scaled time $\tau$ for
$u=1$ and 5, respectively. For larger $u$, $r$ decreases more quickly with $\tau$.
This is not surprising because of the quadratical and cubical dependence on $r$ of
the rates for meson and baryon production. From the fact
that the constituent quark density decreases with $\tau$ very rapidly, we can see
that the average production time for baryons is shorter than that for mesons. Or in other
words, baryons are produced earlier than mesons in the hadronization process.
It has been shown that the pion size measured by interferometry increases with the pion
multiplicity \cite{pis}, steadily with bombarding energies
in similar systems \cite{pis2}. For protons, the measurements of NA49 \cite{na49}
together with measurement at lower energies suggested a very weak dependence of
the two-proton correlation function on bombarding energies. If protons and pions
share the same collective expansion, smaller source size means earlier production.
Our results agree qualitatively with those experimental conclusions.

\begin{figure}[tbph]
\includegraphics[width=0.45\textwidth]{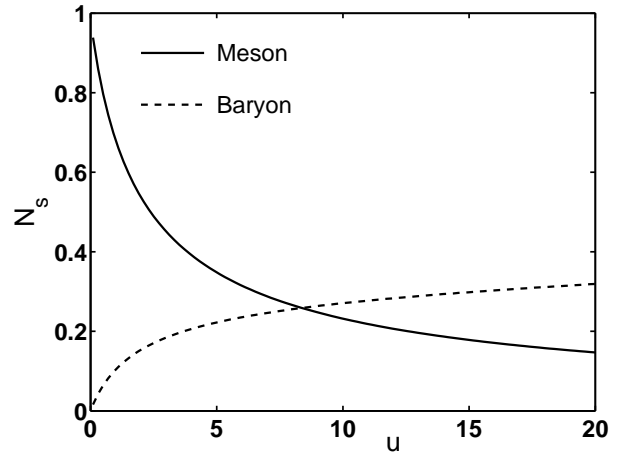}
\caption{Dependence of the scaled multiplicities of mesons and baryons on the scaled
initial density $u$.}
\end{figure}

Once we know the relative quark density as a function of scaled time $\tau$,
one can integrate Eqs. (\ref{nm}) and (\ref{nb}) to get the multiplicities of mesons
and baryons produced in the hadronization process. There exists an unknown constant
factor $VA_1/A_2$. In the following, we define scaled
multiplicities of mesons and baryons as
\begin{equation}
N_s^{M,B}=\frac{3N^{M,B}A_2}{uA_1V}\ .
\end{equation}
One can easily see that $N_s$ is proportional to hadron yield divided by
the quark density just before hadronization. $N_s^{M,B}$ are shown in
Fig. 2 as functions of $u$. As the initial quark density increases, the scaled
yield for mesons decreases, indicating that the meson yield increases slower than
the constituent quark density does. Meanwhile the scaled yield for baryons increase
quickly at small $u$, therefore the baryon yield increases faster than the quark
density. The difference between the trends of $N_M$ and $N_B$ as $u$ increases
can be shown more clearly by the baryon to meson ratio $N^B/N^M$, as in Fig. 3. The ratio
increases with $u$ almost linearly in the range shown. In the near future at LHC,
it can be predicted from Fig. 3 that the baryon to meson ratio will become even larger,
since the initial constituent quark density (or $u$) will be higher. If the
scaled constituent quark density can become high enough (about $u=8$), the yield
of baryons will be about the same as that for mesons, and the baryon to meson ratio
can be more than 2 at $u=20$, if it can be achieved.

\begin{figure}[tbph]
\includegraphics[width=0.45\textwidth]{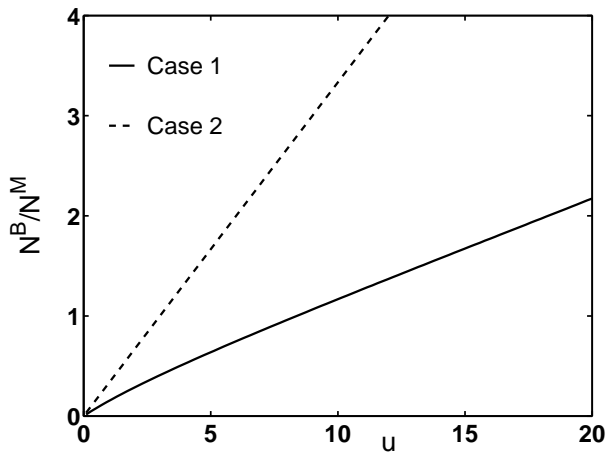}
\caption{Baryon to meson ratio $N^B/N^M$ as a function of scaled density $u$.}
\end{figure}

For the second case we assume that there is no expansion at all. Thus the system
shrinks to 0 from initial volume $V_0$ while the parton density remains the same
in hadronization. Since $\rho$ is now a constant, we get from Eq. (\ref{basic})
\begin{equation}
V\rho=V_0\rho e^{-(A_1\rho+3A_2\rho^2)t}\ .
\end{equation}
In this case the hadronization is much faster than in the first case, because the volume
of parton phase decreases exponentially.
The multiplicities of produced mesons and baryons are
\begin{equation}
N^M=\frac{A_1N_q}{A_1+3A_2\rho}\ ,\ \ \ N^B=\frac{A_2N_q\rho}{A_1+3A_2\rho}\ .
\end{equation}
Then we have the baryon to meson ratio $N^B/N^M=u/3$. Now the ratio turns out to be
proportional to the initial parton density and is larger than the corresponding
values for the first case, as shown in Fig. 3.

In real ultra-relativistic heavy ion collisions the parton system is always expanding
in hadronization. The expansion will slow down the shrinking of the parton volume
in hadronization. Thus the two cases considered in this paper are the two limiting
evolutions of the system. The real dependence
of hadron yield will be in between the results for the two limiting cases.

One may have noticed that we only considered in this paper the contribution to particle
production from the recombination of plenty soft quarks, which is the most important
at low $p_T$ region. This discussion may be applied to particle production at
SPS, RHIC and  LHC energies with different initial parton
density or volume as long as a dense parton phase is created. Thus the rapid increase
of baryon to meson ratio can be a signal of the formation of the dense medium of soft
partons, and the density of the medium can be deduced approximately from the measured
$p/\pi$ ratio. The initial parton density just before hadronization may depend only on the
colliding energy. So the baryon to meson ratio for collisions at fixed center of
mass energy will be proportional to the ratio of the reaction rates, or in other words,
proportional to the ratio of the degeneracy factors, as claimed by the statistical
hadronization model \cite{mun}.

It should be pointed out that the shape of constituent quark distributions can, in reality,
be changing in the hadronization process. As a result, above results may be modified.
However, one can easily imagine that the shape changes more slowly than the normalization
of the distribution. Then the modification will not be too big. In Eqs. (7) and (8) the
shape dependence can be canceled partially in $A_1/A_2$. So, qualitatively, the
results from this investigation will not be changed when the evolution of shapes of
quark distributions is taken into account.

In conclusion, we showed in this paper the importance of the evolution of
constituent quark density in the hadronization process. When the evolution
is taken into account, unitarity can be maintained, and the baryon to meson
ratio is shown to increase monotonically with the initial constituent quark
density.

\acknowledgments{This work was originated when the author was working in the University of
Oregon. He would like to thank Prof. Hwa for the hospitality during his stay
in Eugene and stimulating discussions. Discussions with Prof. Ko from Texas A\&M University
is also acknowledged. This work was supported in part by the National
Natural Science Foundation of China under grant No. 10475032 and by the Ministry
of Education of China under grant No. 03113. }

\end{document}